\documentclass[apj]{emulateapj}
\usepackage{epsfig}
\usepackage{multirow}
\usepackage{graphicx}
\usepackage{epstopdf}
\usepackage{amsmath, amsthm, amssymb}
\usepackage{natbib}
\usepackage{rotating}
\shorttitle{Disk-stability and number of arms}
\shortauthors{D'Onghia}

\begin{document}

\title{Disk-stability constraints on the number of arms in spiral galaxies} 

\author{
Elena D'Onghia\altaffilmark{1}
}
\affil{Department of Astronomy, University of Wisconsin, 475
  N. Charter Street, Madison, WI 53076, USA}
\email{e-mail:edonghia@astro.wisc.edu}

\altaffiltext{1}{Alfred P. Sloan Fellow}

\begin{abstract}
A model based on disk-stability criteria to determine the number of spiral arms 
of a general disk galaxy with an exponential disk, a bulge and a dark halo described by
a Hernquist model is presented. The multifold rotational symmetry of
the spiral structure can be evaluated analytically once the structural properties 
of a galaxy, such as the circular speed curve, and the disk surface
brightness, are known. By changing the disk mass, these models 
are aimed at varying the critical length scale parameter of the disk 
and lead to a different spiral morphology in agreement with prior models. 
Previous studies based on the swing amplification and
disk stability have been applied to constrain the mass-to-light ratio
in disk galaxies. This formalism provides an analytic expression to
estimate the number of arms expected by swing amplification making its
application straight-forward to large surveys.
It can be applied to predict the number of arms in the Milky Way as a function
of radius and to constrain the mass-to-light ratio
in disk galaxies for which photometric and kinematic measurements are
available, like in the DiskMass survey. Hence, the halo contribution to the total mass in the inner parts 
of disk galaxies can be inferred in light of the ongoing and forthcoming
surveys.

\end{abstract}




\section{Introduction}
\label{sec:intro}

A long-standing problem in understanding the dynamics of disk galaxies 
is the uncertainty of the fractional contribution of the dark halo to 
the total mass of the galaxy within the optical radius of the stellar disk.
While galaxies are believed to be dark matter dominated in their outer
parts, at radii much larger than the optical radius,  the assessment of the 
relative fraction of baryons and dark matter in the inner part of disk
galaxies is still debated \citep[e.g.][]{Courteau14}.

The circular-speed curve constrains the total mass distribution at all radii, 
and the disk contribution is usually estimated by combining the light profile 
with an estimate of the mass-to-light ratio of the disk. The difference between the total 
rotation curve and the circular velocity of the disk is assumed to be due to the presence of
dark matter \citep[][]{Bosma78,CarFree85}. Because the mass-to-light ratio of
the disk is uncertain, the relative contributions of the disk and halo
are poorly determined.

Attempts to constrain the mass-to-light ratio come from studies of stellar
populations \citep[e.g.][]{Conroy13},  and comparisons
between dynamical and stellar population mass estimates \citep{BelldeJong01}.
 However, these models depend on assumptions made about the  IMF 
and stellar ages, leading to rather large uncertainties in the disk
contribution.
A maximum disk hypothesis for the mass decomposition has been suggested 
for disk galaxies \citep[][]{vanAlbada85}. It assumes that the total rotation curve of disk galaxies 
out to the optical radius can be fitted by models that assume a mass-to-light
ratio of the disk independent of radius and with no need for dark matter in the inner parts
(maximum disk). A typical maximum disk contributes more than 80\% of the
total circular speed at 2.2 R$_d$, where R$_d$
is the scale length of the exponential disk. 
In terms of disk mass fraction
the maximum disk hypothesis requires a disk contribution of 
$f_d$(2.2 R$_{d}$)=(V$_{disk}$/V$_{tot})^2>$0.7.

Additional arguments have been made to constrain the halo and disk contribution
to the total gravitational field via different techniques. In particular, studies on
Tully-Fisher residuals applied to bright disk galaxies suggest that on average 
disk galaxies are sub-maximal with  $f_d$(2.2 R$_{d}$)=0.25-0.5 \citep{CourteauRix99}.
Another approach assumes that the peak circular velocity
of the stellar disk, measured at R=2.2 $R_d$, is related to the vertical velocity
dispersion, and the scale height through some scaling relations \citep{vanderkruit88}. 
More recently, \citet{Bershady10} applied the velocity dispersion
technique to a sample of 46 nearly face-on galaxies finding that the disk fraction of their sample
ranges between 0.16 and 0.5. A similar method has been recently applied to the Milky Way. The analysis 
includes the ability to disentangle stellar populations and results in a low
estimate of the disk scale length and a correspondingly high disk fraction of
0.69$\pm 0.07$ \citep[][]{BovyRix13}(hereafter BR13).

The study here presents a model based on swing amplification \citep{Toomre81} 
to compute analytically the azimuthal wavenumber at each radius
that is likely to experience the strongest growth to determine
the number of spiral arms expected at that radius\citep{A88,TK91,B99,FA05}. For a sample of 
grand design spirals, \citet{AT87}, hereafter ABP87, applied the swing amplification formalism 
in an attempt to
constrain the mass-to-light ratio of dozens of disks. ABP87 study is of
considerable observational interest, since it treated the disk stars and gas
as distinct components, and computed their  contribution of the total
rotation curve using directly their observed projected surface density
profiles with an appropriate  potential solver. The bulge contribution was estimated in a
similar way. This approach offered the advantage not to rely on a specific disk, bulge
or halo model. It constrained the mass-to-light ratio of disk galaxies and thereby inferred the 
dark matter 
and disk contribution within the optical radius. 
Here the disk-stability criterion is 
generalized for a galaxy characterized by an exponential disk, a
stellar bulge, and a Hernquist halo. It provides 
a simple analytic expression to estimate the number of arms expected by
swing amplification, making its application straight-forward to large surveys. 
  

The model is detailed in Section~\ref{sec:model}. 
Numerical simulations are presented in Section~\ref{sec:simulations}. 
Section~\ref{sec:obse} applies the method 
to an observational sample, and  Section~\ref{sec:conc} summarizes 
the main results.

\section{Model}
\label{sec:model}


We consider the case of a stable disk with  $Q \ge 1$  
responding to gravitational perturbations by swing amplification.    
For a given disk the efficiency of the amplification is characterized by a
factor defined as:

\begin{equation}
X = \frac{\lambda}{\lambda_{crit}}=\frac{\kappa^2}{2\pi G \Sigma}\frac{R}{m}
\end{equation}

\noindent
where the critical wavelength is defined as $\lambda_{crit}$=$4\pi^2 G \Sigma/\kappa^2$, $G$ is the gravitational constant, and 
$\kappa$ is the epicycle frequency.

Numerical studies have shown that for a stable disk, as arms swing from leading to
trailing, the amplification factor 
depends on the value of the $X$ parameter. Particularly suggestive is Fig. 8 of 
\citet{Toomre81} showing that for disks with flat circular velocity curves and $Q=1.2$
the gravitational response to perturbations of a stellar disk is the same as
that of a gaseous
disk and that the swing amplification is strongest when the parameter $X$ is 1.5, 
leading to an amplification of a factor from 40 up to 100
\citep[][]{AT84,Fuchs01}.

Given the circular speed curve, the epicyclic frequency of the galaxy disk is determined.  
For a general galaxy with a bulge, disk and a dark halo, the epicyclic
frequency depends on the total angular frequency, which is the sum of the angular frequencies, according to:

\begin{equation}
\Omega^2=\Omega^2_D+\Omega^2_B+\Omega^2_H
\end{equation}

\noindent
For an exponential disk, the angular frequency is:

\begin{equation}
\Omega^2_D=\frac{G M_D}{2 R^3_h}[I_0(y)K_0(y)-I_1(y)K_1(y)]
\end{equation}

\noindent
where $y=R/2R_d$, $R_d$ is the disk scale length, and $M_D$ is the disk total
mass, including gas and stellar components.
For bulges and halos described by Hernquist models \citep{Hernquist90}, the
angular frequency of the bulge is:

\begin{equation}
\Omega_B^2=\frac{GM_B}{R(R+a_b)^2}
\end{equation}

\noindent
where $a_b$ and $M_B$ are the bulge scale length and mass, respectively, and for the halo:

\begin{equation}
\Omega_H^2=\frac{GM_H}{R(R+a_h)^2}
\end{equation}

\noindent
where $a_h$ and $M_H$ are the halo scale length and the mass, respectively.

The Fourier coefficient, $m$, that characterizes the number of spiral arms, generalized for a galaxy
with those properties becomes:

\begin{align}
m= & \, \frac{e^{2y}}{X} \Big(\Big[\frac{M_B}{M_D}\frac{2y+3a_b/R_d}{(2y+a_b/R_d)^3}\Big] \notag \\
   & \, + \Big[\frac{M_H}{M_D} \frac{2y+3a_h/R_d}{(2y+a_h/R_d)^3} \Big] \notag  \\
   & \, + \frac{y^2}{2}(3I_1K_0-3I_0K_1+I_1K_2-I_2K_1) \notag  \\ 
   & + 4y(I_0K_0-I_1K_1)\Big) 
\end{align}

\begin{figure}[ht]
  \begin{center}
    \includegraphics[width=0.5\textwidth]{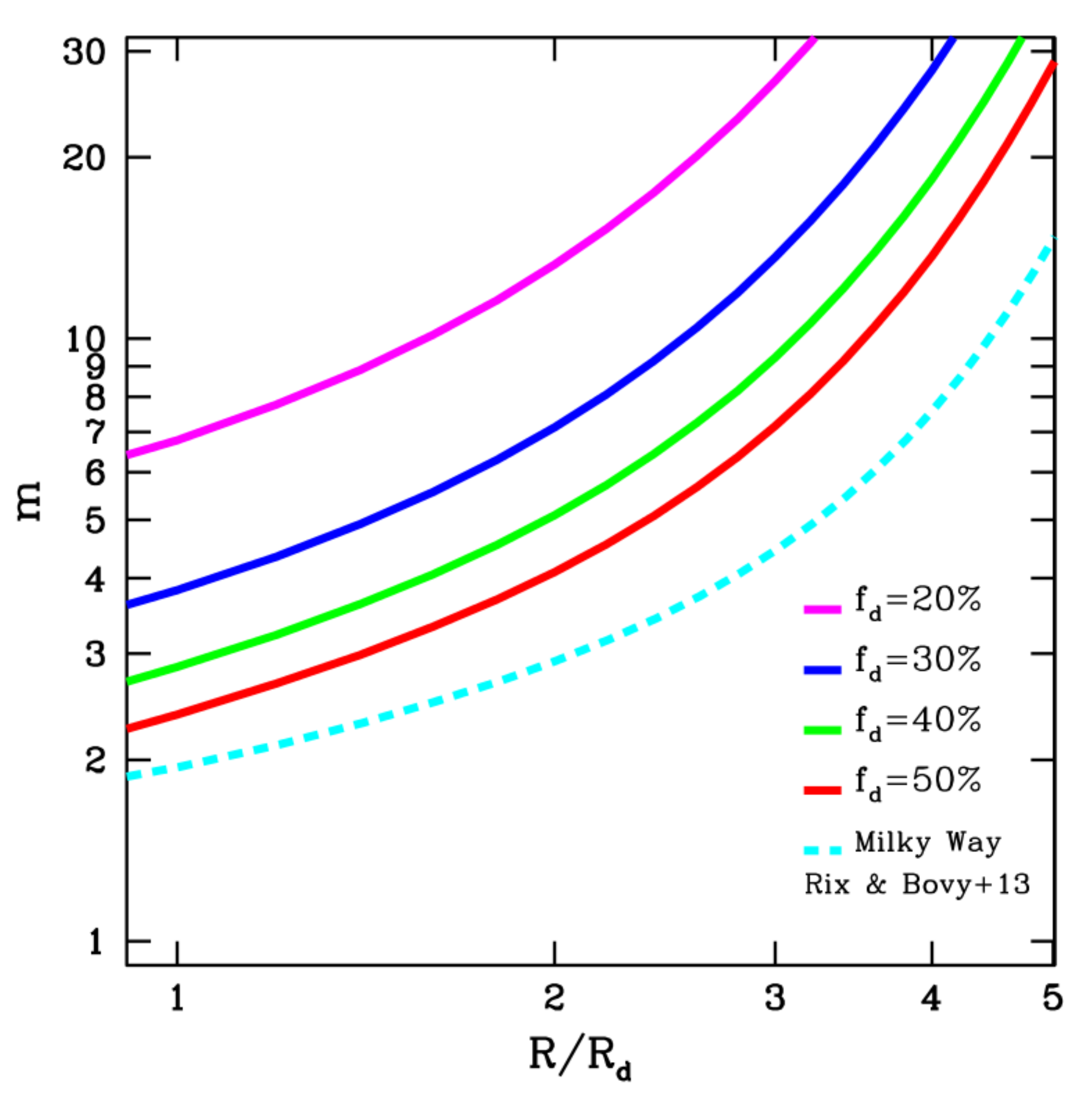}
    \caption{The total number of arms  as a function of the distance
from the center for a galaxy with the halo mass of 9.5$\rm{x}10^{11}$
M$_{\odot}$ and a halo scale length of 29.6 kpc.
Galaxies with the same mass for the halo display a progressive increase of the disk fraction,
$f_d$, from 20\% to 50\%, from the top to bottom. The predicted number of arms
for the Milky Way is also displayed adopting the structural properties of
\citet{BovyRix13} with the inclusion of a bulge with a mass of 4x$10^9$ M$_{\odot}$ and a
  scale radius of 600 pc. The assumed amplification factor is $X$=1.5.}
\end{center}
\label{fig:nspirals}
\end{figure}

\noindent
For a galaxy with a halo mass of 9.5$\rm{x} 10^{11}$ M$_{\odot}$ the total
number of arms $m$ as a function of the distance
from the galaxy center  is displayed in Fig. 1.
The different lines display models where the total mass distribution of the
dark halo is kept fixed but the self-gravity of the disk is progressively
increased. In particular, the disk mass fraction, $f_d$, within 2.2 scale
lengths, rises from 20\% to 50\%, from the top to bottom. The disk scale
lengths assumed are the same as in the numerical simulations presented in the
next section. No bulge is assumed  for simplicity except for the Milky
Way model where parameters are assumed from BR13. 
 
Note that the total number of arms is expected to increase with distance
from the galaxy center as the highest Fourier components $m$, which are always present,
dominate the spiral structures in the outer
parts of the disk. This is in agreement with previous models and
  comparison with observations of ABP87. 
However, exponential disks are less
self-gravitating as we move further out from the center, thus the arms in the outer parts of
the disk will have lower strength and may just be in the form of poorly
amplified wakelets \citep[][]{Donghia13}.

Fig. 1 shows that at a given radius, the total number of spiral arms decreases as the disk fraction 
increases, as expected. The predicted number of arms of the Milky Way is also included
in Fig. 1 (dashed line) adopting the structural parameters presented in BR13, with the disk fraction being estimated to  be $f_d \approx$0.69 within 2.2
$R_d$. Note that according to this analysis the Milky Way should have two 
spiral arms at approximately  4.50 kpc, nearly at the edge of the stellar bar, and 
a total of 5-6 spiral arms, lower in strength, in the solar neighborhood.

\section{Numerical experiments}
\label{sec:simulations}

\noindent
Numerical simulations of various disk models have shown that changes to the disk 
mass lead to varying the critical length scale
parameter ${\lambda}_{crit}$ and lead to a
different spiral morphology (see e.g. \citet{SC84,Carlberg85,NDH12}). 

Here a set of numerical simulations are employed to determine the extent to which the response of the
disk depends specifically on the mass distribution of the galaxy and on the
disk self-gravity. 

In order to test the formalism introduced above, a set of numerical simulations 
are carried out with the parallel TreePM code GADGET-3 (last described
in \citet{Springel05}). Following \citet{Donghia13},  the tree-based gravity solver 
is employed coupled 
with a static potential to solve for the evolution of collisionless particles.
The galaxies in this study consist of dark matter halos and rotationally supported
stellar disks of 10$^7$ particles. The parameters describing each component are independent 
and the models are constructed similarly to the approach described
in \citet{Hernquist93}. The dark matter mass distribution is modeled assuming a Hernquist profile with
a total halo mass and a scale length of 9.5$\rm{x} 10^{11}$ M$_{\odot}$ and 29.6 kpc, respectively. 
The stellar disk is modeled according to the initial conditions as a thin exponential surface density profile
of scale length $R_d$, so that the stellar disk mass is $M_*=f_d M_{tot}$ where $f_d$ is the disk fraction of the total mass of the galaxy. The vertical mass distribution of the stars in the disk is specified by the profile of an
isothermal sheet with a radially constant scale height $z_0$.
A value of $z_0=0.1$ R$_d$ is adopted.

In a sequence of experiments shown in Fig. 2, the contribution of the
disk to the total mass of the galaxy is altered. Like
in Fig. 1 the total mass of the halo is fixed but the disk mass is increased
in order to obtain a disk fraction, $f_d$, within 2.2 disk scale lengths 
varying from 20\% to 50\% of
the total mass. Disk scale lengths are set in the initial conditions to values
that conserve the total angular momentum of the disk. They range from 2.75 kpc
in the model with $f_d$=0.5 to 3.3 kpc in the one with $f_d$=0.2. 
Fig.2 shows that the galaxy ranges from the multi-armed
spirals obtained for disks with low mass fractions (lower panels in Fig. 2), to those with  four
prominent arms typical of galaxies where the disk contributes about 50\% of
the total mass (top left panel in Fig. 2).
As anticipated by the analytic arguments and previous studies, the results 
shown in Fig. 2 confirm that the spiral 
morphology in the simulated disks is indeed determined by the
structural properties of the galaxy through the expression of $\lambda_{crit}$.

\begin{figure}
  \begin{center}
    \includegraphics[width=0.5\textwidth]{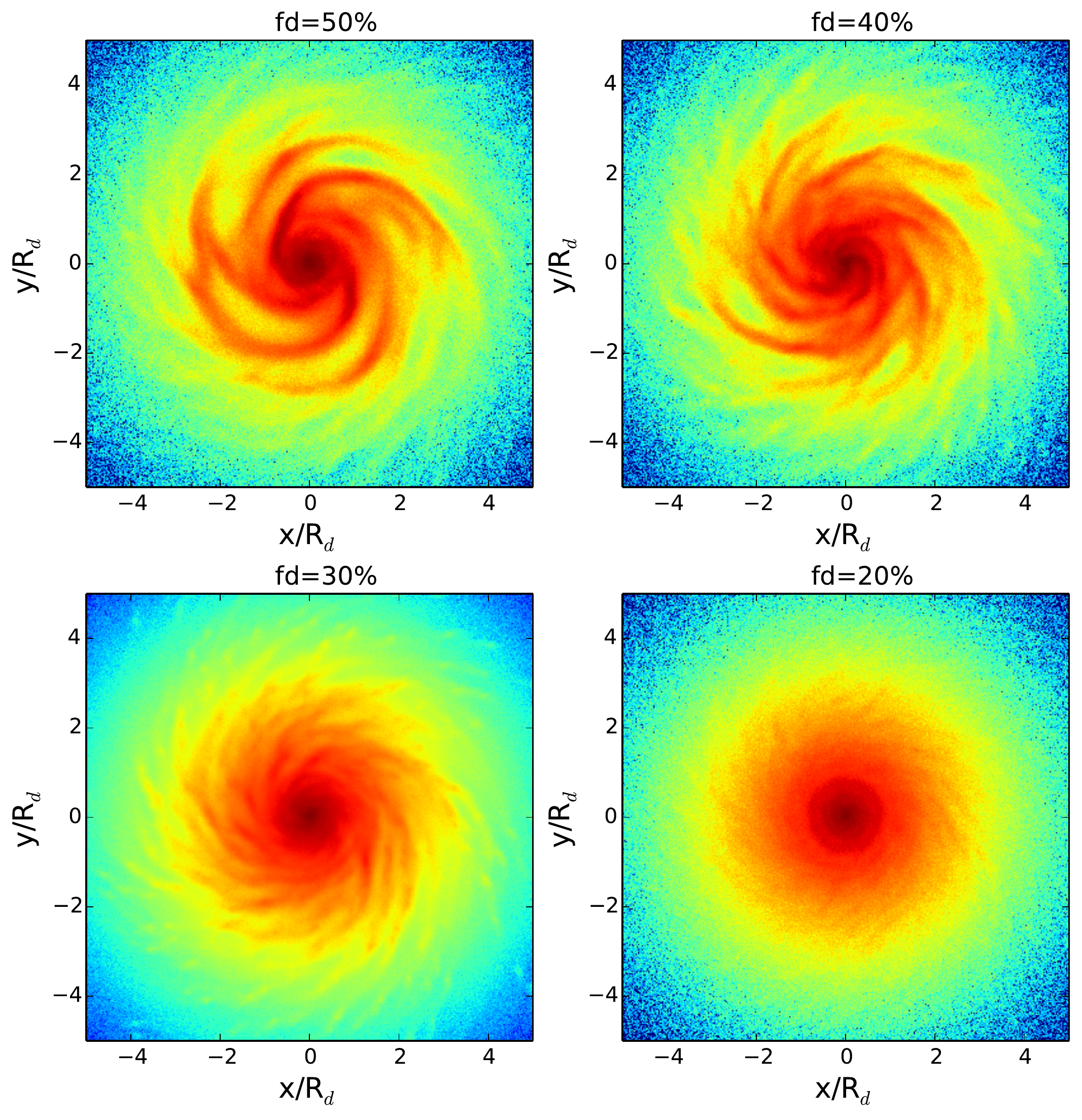}
  \caption{Spiral morphologies of the simulated stellar disk displayed face-on
    after one galactic year. The disk mass fraction at 2.2 scale lengths, $f_d$ ranges from 50\% of the
total mass (top left panel) to 20\% (bottom right panel).}
\label{fig:image}
\end{center}
\end{figure}
A quantitative analysis of the strength of the arms as a
function of the galaxy radius has been performed through a Fourier analysis of the surface mass density of
the stellar disk. The results show that the amplitude of the arms, on average, ranges from
10\% of the stellar background for the
multi-armed galaxies (right bottom panel of Fig. 2)  to 50\% for the galaxies
with four arms and approximately four-fold rotational symmetry (top left panel
in Fig. 2). The number of arms increases as a function of radius in agreement
with the analytic estimates.

\section{Application to the {\it DiskMass} Sample}
\label{sec:obse}
\noindent
The formalism introduced can then be applied to the Milky Way and to the DiskMass sample,
consisting  of  
face-on galaxies with rotation velocities between 100 kms$^{-1}$ and 250 kms$^{-1}$  \citep{Bershady10,Davis15}. 
Models are run assuming the structural properties of the galaxies, when
available, in particular scale lengths, and total mass as  reported by
\citet{Martinsson13}, with the observed disk fraction $f_d$ ranging from 0.16
to 0.5. As expected, Fig. 3 shows a correlation between the total number of spiral
arms predicted by the swing amplification and the disk fraction. Submaximal disks are thus expected to be 
multi-armed galaxies, and the two-armed galaxies are the ones 
with higher $f_d$ within 2.2 scale lengths.
\begin{figure}
  \begin{center}
    \includegraphics[width=0.49\textwidth]{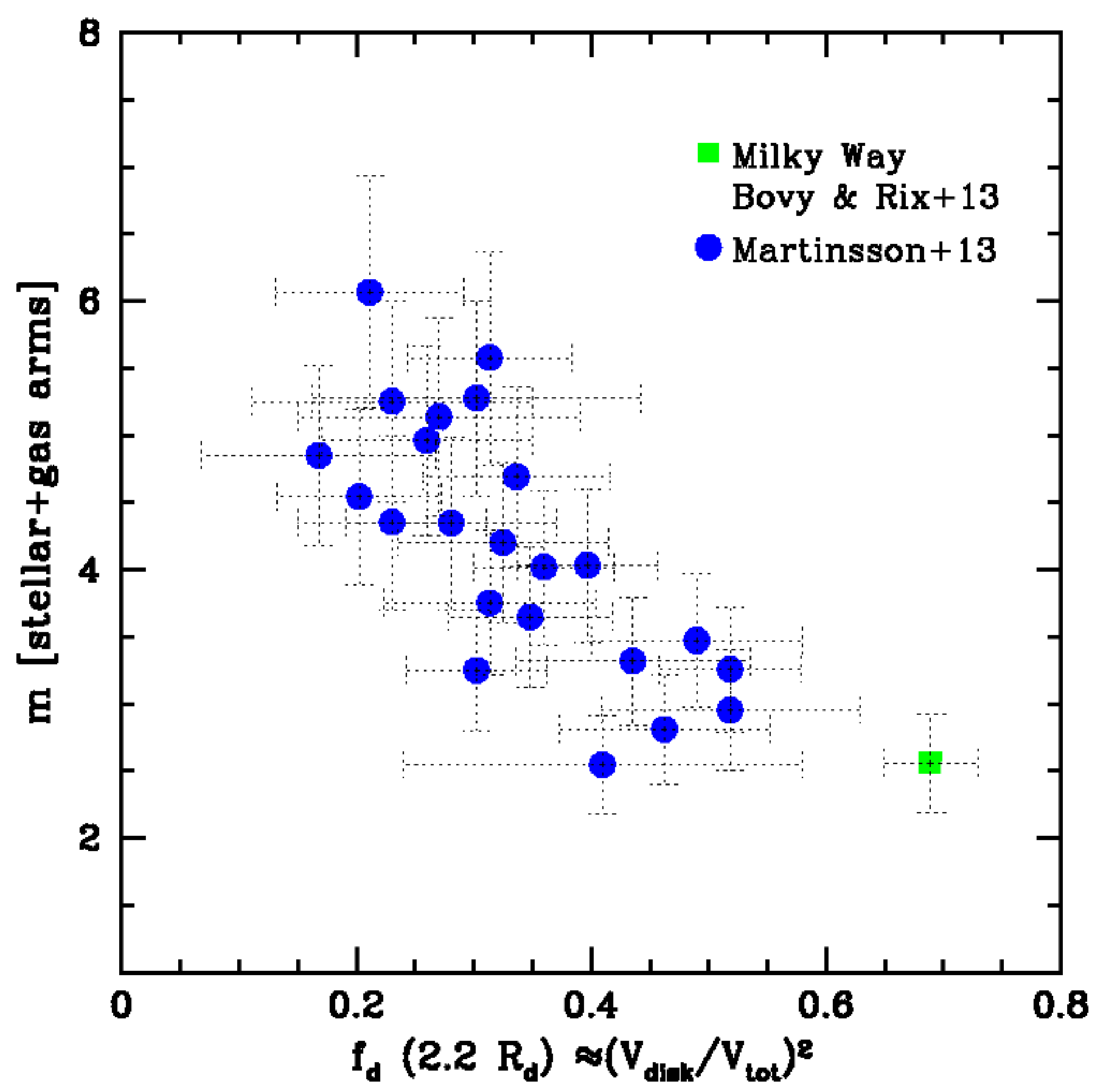}
  \caption{Total number of spiral arms predicted by models applied to the
    DiskMass catalog of galaxies (blue symbols) as a function of the disk
fraction, $f_d$, measured at 2.2 disk scale lengths. $m$ is the average value
estimated by models assuming 1.5 $\le X \le$2 with the error
bars given by the two extreme values. The prediction of the number of spiral arms of the Milky Way is also included
(green symbol) for the structural parameters of the Milky Way reported in \citet{BovyRix13}.}   
\end{center}
\end{figure}
A visual inspection of the images of the galaxy sample in the available bands 
supports this result, although a more quantitative analysis is needed.

\section{Conclusions}
\label{sec:conc}

A model based on the swing amplification mechanism is generalized for
a realistic galaxy model with an exponential stellar disk and a dark halo
described by the Hernquist mass profile. The number of spiral arms is derived 
in the context of disk stability.
More generally, the model confirms previous results that show that the number of arms in galaxies should 
depend only on the structural properties of the galaxy, with a correlation
between the disk fraction and the dominant wave mode numbers. 
These models are aimed at providing an additional constraint on the number of spiral arms in the Milky
Way as a function of the radius and an independent test on the structural parameters of the Milky Way
provided by BR13 (as shown in Fig.3). 
They have been applied to the DiskMass survey to confirm the known correlation 
between the total number of spiral
arms predicted by the swing amplification theory and the disk mass fraction. 

Ongoing and forthcoming surveys of disk galaxies provide strong tests of
the applicability of these models. In particular, they can constrain the mass-to-light ratio of disk galaxies and hence to infer the dark matter 
and disk contribution within the optical radius. Although the study presented
in ABP87 is model independent and with a more accurate treatment of the swing
amplification, the formalism presented here has the advantage to provide an
analytic expression for the number of arms expected by swing amplification
that can be used to compare with the actual number of arms observed in large
sample of galaxies. For a given 
rotation curve the epicycle frequency is determined analytically. If the number of arms
expressed by $m$ at a given radius can be measured independently,
that number indicates the azimuthal wavenumber undergoing the strongest 
swing amplification at that radius (according to eq.(6) for $X=1.5-2$ for any
given galaxy). Therefore the surface density of the disk can be derived analytically from eq.(1),
leading to an independent constraint in the mass-to-light ratio of the disk, and hence to the halo
contribution to the total mass.

\vspace{0.5cm}

I am especially grateful to A. Quillen, L. Hernquist, S. Sessarego for much
discussion and encouragement. I gratefully acknowledge the anonymous referee and the 
support of the Alfred P. Sloan Foundation and the hospitality of the Aspen
Center for Physics, funded by NSF Grant No. PHY-1066293.
This work is funded by NSF Grant No. AST-1211258 and ATP-NASA Grant No. NNX14AP53G.  



\end{document}